\documentclass[aps,prb,onecolumn,groupedaddress, showpacs]{revtex4-1}

\usepackage{graphicx,amssymb, amsmath}
\usepackage{datetime}
\usepackage{xr}

\newcommand{\dbar}[1]{\bar{\bar{#1}}}
\newcommand{\com}{{\scriptscriptstyle{\mathbb{C}}}}
\newcommand{\rmi}{{\rm i}}
\newcommand{\ff}{{\it ff}}
\newcommand{\nf}{{\it nf}}
\renewcommand{\i}{{\it (i)}}
\newcommand{\eff}{{\it eff}}
\newcommand{\xma}[1]{{\dbar{#1}_{\scriptscriptstyle\times}}}
\let\originalleft\left
\let\originalright\right
\renewcommand{\left}{\mathopen{}\mathclose\bgroup\originalleft}
\renewcommand{\right}{\aftergroup\egroup\originalright}
\newcommand{\parallelslant}{\mathbin{\scriptscriptstyle \!/\mkern-5mu/\!}}

\newcommand{\JLTP}{J. Low Temp. Phys. }
\newcommand{\JPCSSP} {J. Phys. C: Solid State Phys. }
\newcommand{\PC}{Physica C }
\newcommand{\JSNM}{J. Supercond. Nov. Magn. }
\newcommand{\JS}{J. Supercond. }
\newcommand{\PRB}{Phys. Rev. B }
\newcommand{\PRL}{Phys. Rev. Lett. }
\newcommand{\JAP}{J. Appl. Phys. }
\newcommand{\SUST}{Supercond. Sci. Technol. }
\newcommand{\JPSJ}{J. Phys. Soc. Jpn. }
\newcommand{\NatPhys}{Nat. Phys. }
\newcommand{\ZPB}{Z. Phys. B }
\newcommand{\EPL}{Europhys. Lett. }
\newcommand{\JETP}{Soviet Phys. JETP }
\newcommand{\APL}{Appl. Phys. Lett. }
\newcommand{\EPJB}{Eur. Phys. J. B }
\newcommand{\RPP}{Rep. Prog. Phys. }
\newcommand{\RSI}{Rev. Sci. Instrum. }
\newcommand{\RMP}{Rev. Mod. Phys. }
\newcommand{\MST}{Meas. Sci. Technol.}
\newcommand{\IJMPB}{Int. J. Mod. Phys. B }
\newcommand{\IEEEas}{IEEE Trans. Appl. Supercond. }
\newcommand{\IEEEmag}{IEEE Trans. on Magnetics }

\begin{document}

\title{Theory of measurements of electrodynamic properties in anisotropic superconductors in tilted magnetic fields. Part II: high frequency regimes.}%

\author{N. Pompeo}%
\email{pompeo@fis.uniroma3.it}
\affiliation{Dipartimento di Fisica ``E. Amaldi'' and Unit\`a CNISM, Universit\`a Roma Tre, Via della Vasca Navale 84, 00195 Roma, Italy}%

\date{November 15th, 2012}%

\begin{abstract}
The model for high frequency electrodynamics in anisotropic type-II superconductors in the vortex state is studied considering arbitrary orientations between the applied field, the applied current and the anisotropy axis. An anisotropic treatment is provided for the vortex dynamics, taking into account all the phenomena relevant at high frequency, which include flux flow, pinning and creep. The coupling between vortex motion and high frequency currents is included, providing an entirely tensor model of the electromagnetic response to high frequency fields. 
Examples of data analysis of angular measurements are presented, showing how to derive the angular dependence of the material properties from the measured anisotropic response. Finally, the expression of the measured angle-dependent surface impedance in the largely used thin film geometry is computed.
\end{abstract}


\maketitle

\section{Introduction}

Material anisotropy, which characterizes many superconductors of wide interest,\cite{anisotropicSC} has a profound impact, among the others, on their vortex dynamics and on the related pinning phenomena. 

In this paper I focus on the high frequency regime of the electrical transport properties in the mixed state, largely studied because of the great deal of information that they can provide.

Due to the anisotropy of the superconductor, the transport properties depend on the various angles between the anisotropy axes and the applied field and current, so that the measured quantities have a non-straightforward relationship with the material properties.\cite{Pompeo_Part1}

While some aspects of the problem in the d.c. and (low frequency) a.c. regimes have been addressed in previous works,\cite{previousWorks, haoclem, CCaniso, coffeyJLTP} I proposed\cite{Pompeo_Part1}  a generalized treatment centered on the force equation for the vortex motion in the linear regime, including both material anisotropy and pinning, in uniaxial anisotropic superconductors in a magnetic field applied with generic orientation. The very different free flux flow regime (dominated by dissipation) and pinned Campbell regime (dominated by pinning) were addressed. The tensor expressions for the resistivity and for vortex parameters like the viscous drag, the vortex mobility and the pinning constant were given, and the measurable quantities for arbitrary angles between magnetic field, current and the anisotropy axis were derived.

In this work I extend the treatment of Ref. \onlinecite{Pompeo_Part1} to the high frequency regime, where additional phenomena emerge. Indeed, by increasing the frequency, dissipation and pinning effects become comparable, requiring to be simultaneously taken into account. Moreover, even within the limit of small currents, vortex creep effects become relevant.\cite{SongPRB79, APL91, creepCorbino, BrandtModel,CCiso} Finally, the coupling between vortex motion and the high frequency currents (due to both superfluid and quasiparticles) contribute significantly.\cite{}

This work is organized as follows: first, I generalize the anisotropic vortex dynamics model to include both dissipation and pinning effects, the latter including the effects of thermal depinning/creep (Sec. \ref{sec:rhovmtensor}). Contextually I provide examples of experimental data analysis (Sec. \ref{sec:rhovm_examples}). Second, I consider the full coupling between vortex motion and the high frequency currents (Sec. \ref{sec:CCmodel}). Finally, I provide an application example of the full model by computing the expression of the measured surface impedance in the largely used thin film geometry (Sec. \ref{sec:Zfilmaniso}).

\section{A.c. vortex motion resistivity}
\label{sec:rhovmtensor}

In this Section I address the issue of the vortex motion resistivity tensor $\dbar{\rho}_v$ in a uniaxial superconductor, and I calculate explicit expressions for the angular dependence, with arbitrary angle between the applied field $\vec{B}$, the applied current density $\vec{J}$ and the crystal axis.  
In order to introduce all the relevant quantities, I first recall the general models describing the high frequency vortex dynamics with reference to the commonly studied configuration of isotropic superconductors with $\vec{J}\perp\hat{B}$ (Sec. \ref{sec:scalarmodel}).
Second, I develop the full treatment for uniaxially anisotropic superconductors (Sec. \ref{sec:acforceequation}). Finally, some examples of data analysis based on the obtained results will be illustrated (Sec. \ref{sec:rhovm_examples}).

\subsection{Short review of scalar models}
\label{sec:scalarmodel}

The scalar force equation describing the vortex motion in an isotropic superconductor with isotropic (point) pinning and $\vec{J}\perp\hat{B}$ is, in the sinusoidal regime $e^{\rmi\omega t}$:\cite{GR,Golo}
\begin{equation}
\label{eq:forceIso}
\eta v+\frac{k_p}{\rmi\omega}v=\Phi_0 J+F_{therm}
\end{equation}
where $v$ is the vortex velocity, $\eta$ is the vortex viscosity, $k_p$ is the pinning constant (also called the Labusch parameter) yielding an elastic recall force $-k_p u$, with $u= v/(\rmi \omega)$ the vortex displacement from the pinning center, and $F_{therm}$ is a stochastic thermal force causing thermal depinning (vortex creep). 
Different approaches,\cite{BrandtModel,CCiso} with different ranges of applicability,\cite{PompeoPRB} have been proposed to take into account creep effects. 
As an illustration, I follow here the description of thermal depinning in terms of the relaxation of the pinning constant $k_p(t)=k_p e^{-t/\tau_{th}}$ (Ref. \onlinecite{BrandtModel}). The characteristic time for thermal activated depinning is:
\begin{equation}
\label{eq:tauth}
\tau_{th}=\tau_p e^{U/K_BT}
\end{equation}
where $\tau_p$ is the inverse of the (de)pinning angular frequency $\tau_p^{-1}=\omega_p=\eta/k_p$ (which will be commented on later) and $U$ is the activation energy. Equation \eqref{eq:forceIso} can then be rewritten as:\cite{BrandtModel}
\begin{equation}
\label{eq:forceIso1}
\eta v+\frac{k_p}{\rmi\omega}\frac{1}{1-\frac{\rmi}{\omega\tau_{th}}}v=\eta_\com v=\Phi_0 J
\end{equation}
where the complex viscosity $\eta_\com$ has been introduced:
\begin{equation}
\label{eq:etac0}
\eta_\com=\eta\left(1-\rmi\frac{\omega_p}{\omega}\frac{1}{1-\frac{\rmi}{\omega\tau_{th}}}\right)
\end{equation}
The corresponding scalar vortex motion resistivity is:
\begin{equation}
\label{eq:rhoBiso}
\rho_{v}=\frac{\Phi_0 B}{\eta_\com}=\frac{\Phi_0 B}{\eta}\frac{\varepsilon'+\rmi\frac{\omega}{\omega_0}}{1+\rmi\frac{\omega}{\omega_0}}
\end{equation}
where the characteristic angular frequency $\omega_0$ is:
\begin{equation}
\label{eq:omega0}
\omega_0=\tau_{th}^{-1}+\tau_p^{-1}
\end{equation}
and the creep factor $\varepsilon'$ is:
\begin{equation}
\label{eq:creep}
\varepsilon'=\frac{1}{1+e^\frac{U}{K_BT}}
\end{equation}
For $U\rightarrow\infty$ the creep is negligible, $\varepsilon'\rightarrow0$ and $\omega_0\rightarrow\omega_p$; consequently the vortex motion resistivity becomes: 
\begin{equation}
\label{eq:rhoGRiso}
\rho_{v}=\frac{\Phi_0 B}{\eta}\frac{1}{1-\rmi\frac{\omega_p}{\omega}}=\left(\rho_\ff^{-1}-\rmi\rho_C^{-1}\right)^{-1}
\end{equation}
where $\rho_\ff=\Phi_0 B/\eta$ is the flux flow resistivity and $\rho_C={\Phi_0 B}/({k_p}\omega)$  is the well-known Campbell resistivity.\cite{Campbellpenetration, Pompeo_Part1}

This limit corresponds to the Gittleman\textendash{}Rosenblum (GR) model.\cite{GR} From Eq. \eqref{eq:rhoGRiso} it can be seen that the pinning angular frequency $\omega_p$ marks the transition between a ``low frequency'' and a ``high frequency'' regime: for $\omega\ll\omega_p$ the pinning force dominates over the viscous drag, yielding $\rho_{v}\rightarrow\rmi\rho_C$, while for $\omega\gg\omega_p$, a purely dissipative flux flow regime is recovered with $\rho_{v}\approx\rho_\ff$ yielding the same behaviour as in d.c. with no pinning.

Before concluding this short review, it is worth recalling the definition of the often used\cite{rparameter1,tsuchiya,rparameter2} dimensionless ratio $r$:
\begin{equation}
\label{eq:r}
r=\frac{\Im(\rho_{v})}{\Re(\rho_{v})}
\end{equation}
which, if creep is negligible (GR limit) yields:
\begin{equation}
\label{eq:rGR}
r=\frac{\omega_p}{\omega}=\frac{\rho_\ff}{\rho_C}
\end{equation}
The $r$ parameter can be directly computed from the complex resistivity $\rho_v$ and it is unaffected by any systematic error in the experiments. Physically, it allows to easily evaluate whether the vortex dynamics is in the pinning dominated ($r\gg1$) or flux flow dominated ($r\ll1$) regime.

\subsection{The anisotropic model}
\label{sec:acforceequation}

\begin{figure}[ht]
\centerline{\includegraphics[width=4cm]{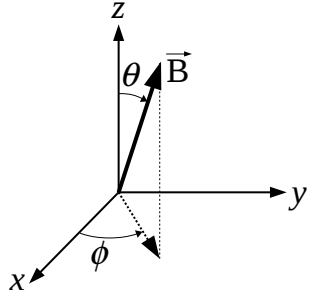}}
  \caption{ Principal frame of reference. The magnetic induction field $\vec{B}$ is also depicted, applied along a generic direction at the polar $\theta$ and azimuthal $\phi$ angles.}
\label{fig:ref}
\end{figure}

In this Subsection I introduce the anisotropy following the approach previously used.\cite{kogan} All symbols will follow the notation introduced in Part I.\cite{Pompeo_Part1} For the sake of completeness, I recall here some basic features. The crystallographic axes are taken as  $x\equiv a$, $y\equiv b$ and $z\equiv c$. The latter is the axis of anisotropy. Figure \ref{fig:ref} shows the chosen frame of reference and the magnetic field $\vec{B}=B\widehat{B}=B(\sin\theta\sin\phi,\sin\theta\cos\phi,\cos\theta)=(B_x,B_y,B_z)$ with generic orientation. 
The anisotropy is considered through the mass tensor:\cite{kogan}

\begin{equation}
\label{eq:masstensor}
\begin{pmatrix}
  m_{ab} & 0 & 0 \\
  0 & m_{ab} & 0 \\
  0 & 0 & m_c
\end{pmatrix}
=
m_{ab}
\begin{pmatrix}
  1 & 0 & 0 \\
  0 & 1 & 0 \\
  0 & 0 & \gamma^2
\end{pmatrix}
=
m\dbar{M}
\end{equation}

\noindent having defined the in-plane mass $m_{ab}=m$, the out-of-plane mass $m_c$ and the anisotropy factor $\gamma^2=m_c/m_{ab}$. The notations $\dbar{A}$, $\vec{A}$ and $\widehat{A}$ denote a tensor/matrix, a vector and a unit vector, respectively.
A static magnetic field is uniformly applied to the superconductor with generic orientation. Vortices are considered as straight and rigid lines, so that more complex phenomena such as helical instabilities of the vortex lines\cite{brandtone} are not taken into account. I assume the validity of the London approximation $B\simeq\mu_0 H$.

The force equation in anisotropic superconductors including both viscous drag and pinning, momentarily neglecting vortex creep (which will be treated later), can be written as:
\begin{equation}
\label{eq:force3a}
\dbar{\eta}\vec{v}+\frac{1}{\rmi\omega}\dbar{k}_p\vec{v}=\Phi_0 \vec{J}\times\hat{B}
\end{equation}
which, by defining the complex viscosity tensor as:
\begin{equation}
\label{eq:etac1}
\dbar{\eta}_{\com}=\dbar{\eta}-\rmi\frac{\dbar{k}_p}{\omega}
\end{equation}
can be also written as:
\begin{equation}
\label{eq:force3b}
\dbar{\eta}_\com\vec{v}=\Phi_0 \vec{J}\times\hat{B}
\end{equation}

It is evident that the force equation \eqref{eq:force3b} is formally equivalent to force equations written for the flux flow and Campbell regimes in Ref. \onlinecite{Pompeo_Part1}. Hence, it is straightforward to apply the electrodynamics model developed there to the series of quantities $[\dbar{\eta}_\com, \dbar{\eta}^\i_\com, \dbar{\rho}_{v}^\i, \dbar{\sigma}_{v}^\i, \dbar{\mu}_v, \dbar{\rho}_{v}]$ which are equivalent to the series $[\dbar{\eta}, \dbar{\eta}^\i, \dbar{\rho}_\ff^\i, \dbar{\sigma}_\ff^\i, \dbar{\mu}_v, \dbar{\rho}_\ff]$ and $[\dbar{k}_p^\i/(\rmi\omega), \rmi\dbar{\rho}_C^\i, \rmi\dbar{\sigma}_C^\i, \dbar{\mu}_v, \rmi\dbar{\rho}_C]$, previously extensively studied.\cite{Pompeo_Part1} It should be recalled that an essential feature of the anisotropic electrodynamics resides in the difference between measurable properties and material properties. As an example, the measured resistivity 
tensor $\dbar{\rho}$ is different from the intrinsic resistivity tensor $\dbar{\rho}^\i$. Here and henceforth, the superscript ``(i)'' indicates material, intrinsic properties.

As a first result, within this framework one can write:\cite{Pompeo_Part1}
\begin{equation}
\label{eq:etac2}
\dbar{\eta}_{\com}=-\xma{B}\dbar{\eta}_{\com}^\i\xma{B}
\end{equation}
where (see Appendix of Ref. \onlinecite{Pompeo_Part1}):
\begin{equation}
\label{eq:crossproduct}
\xma{B}:=
\begin{pmatrix}
  0 & -B_{z} & B_{y} \\
  B_{z} & 0 & -B_{x} \\
  -B_{y} & B_{x} & 0
\end{pmatrix}
\nonumber
\end{equation}
Equation \eqref{eq:etac2} and \eqref{eq:etac1} allow to write the ``intrinsic'' tensor:
\begin{equation}
\label{eq:etaci1}
\dbar{\eta}_{\com}^\i=\dbar{\eta}^\i-\rmi\frac{\dbar{k}_p^\i}{\omega}\\
\end{equation}
In Ref. \onlinecite{Pompeo_Part1} I obtained, neglecting Hall contributions and for weak random point pinning:
\begin{subequations}
\label{eq:etakpi}
\begin{align}
\dbar{\eta}^\i(B, \theta)&=\eta^\i_{11}(B/B_{c2}(\theta))\dbar{M}^{-1} \\
\label{eq:kp1}
\dbar{k}_{p}^\i(B, \theta)&=k_{p,11}^\i(B/B_{c2}(\theta))\dbar{M}^{-1}
\end{align}
\end{subequations}
where the relation $\eta_{ii}^\i=\Phi_0B/\rho_{ii}^\i$ connects the viscous drag and the flux flow resistivity intrinsic tensors. 
From Eqs. \eqref{eq:etakpi}, it is seen that $\dbar{\eta}^\i$ and $\dbar{k}_{p}^\i$ obey the so-called angular scaling law\cite{BGL, blatterone,haoclem}, according to which, in the London approximation, a thermodynamic or intrinsic transport property of a uniaxially anisotropic superconductor depends on the applied field magnitude $B$ and directional angle $\theta$ only through the ratio $B/B_{c2}(\theta)$.
Substituting Eqs. \eqref{eq:etakpi} into \eqref{eq:etaci1} yields:
\begin{align}
\label{eq:etaci2} 
\nonumber
\!\dbar{\eta}_{\com}^\i\!(B, \theta)\!&=\!\left(\!\eta^\i_{11}(B/B_{c2}(\theta))\!-\!\rmi \frac{k_{p,11}^\i(B/B_{c2}(\theta))}{\omega}\!\right)\!\dbar{M}^{-1}\!=\\ 
&=\left(1-\rmi\frac{\omega_p(B/B_{c2}(\theta))}{\omega}\right)\dbar{\eta}^\i(B/B_{c2}(\theta))
\end{align}
It is evident that $\dbar{\eta}_{\com}^\i$ inherits the angular dependencies and anisotropic properties from $\dbar{\eta}^\i$ and $\dbar{k}_{p}^\i$, represented by $\dbar{M}$ and by the scaling law. 

I stress that, in obtaining Eq. \eqref{eq:etaci2}, I have used a very important result for the analysis of the experiment, namely:
\begin{equation}
\label{eq:omegap}
\frac{k_{p,ii}^\i(B/B_{c2}(\theta))}{\eta^\i_{ii}(B/B_{c2}(\theta))}\!=\!\omega_{p,ii}(B/B_{c2}(\theta))\!=\!\omega_p(B/B_{c2}(\theta))
\end{equation}
which holds for $i=1..3$, i.e. for all the principal axes directions. The above equation shows that, contrary to the viscosity or the pinning constant, the pinning frequency is a scalar.\cite{Note1} This fact implies that, whichever orientation is set for $\vec{B}$ and $\vec{J}$, the vortex system will always have the same pinning frequency at fixed $B/B_{c2}(\theta)$. This property suggest a straightforward method to check whether directional defects influence vortex motion: by plotting $\omega_p$ vs the scaled field $B/B_{c2}(\theta)$, any deviation from a scaling curve should indicate the influence of some directional effect, other than the material anisotropy.
Obviously, if extended defects are present as sources of pinning, Eq. \eqref{eq:kp1} does not hold and the above no longer applies.

The intrinsic conductivity and resistivity tensors are:\cite{Pompeo_Part1}
\begin{equation}
\label{eq:rho_vi1}
\dbar{\rho}_v^\i=\left(\dbar{\sigma}_{v}^{\i}\right)^{-1}=\Phi_0B \left(\dbar{\eta}_{\com}^{\i}\right)^{-1}
\end{equation}
Using Eq. \eqref{eq:etaci2}, the explicit expression for $\dbar{\rho}_v^\i$ can be written down:
\begin{equation}
\label{eq:rho_vi2}
\dbar{\rho}_v^\i(B,\theta)\!\!=\!\frac{\rho_{\ff,11}^\i\!(B/B_{c2}(\theta))}{1\!-\rmi\frac{\omega_{p}(B/B_{c2}(\theta))}{\omega} }\dbar{M}\!\!=\!\rho_{v,11}^\i\!(B/B_{c2}(\theta))\dbar{M}\!
\end{equation}
This is a second important results of this work: similarly to $\dbar{\eta}_{\com}^\i$,  $\dbar{\rho}_{v}^\i$ retains the same anisotropy of the flux flow and pinning tensors, given by the mass anisotropy tensor $\dbar{M}$ alone. Moreover, it satisfies the angular scaling law (with a BGL scaling factor\cite{BGL} $s_\rho=1$, as \cite{BGL, Pompeo_Part1} $\rho_\ff$ and $\rho_C$). It is important to remember that $\dbar{\rho}_{v}^\i$ of Eq. \eqref{eq:rho_vi2} is an intrinsic quantity, and as such not necessarily it is directly measurable. 
The actually measurable tensor, according to the results developed for the flux flow regime in Ref. \onlinecite{Pompeo_Part1}, is:
\begin{equation}
\label{eq:rho_v}
\dbar{\rho}_{v}(B,\theta,\phi)=\rho_{v,11}^\i(B/B_{c2}(\theta))\left[\frac{\dbar{\mathcal{M}}(\theta,\phi)}{\epsilon^2(\theta)}\right]
\end{equation}
with $\dbar{\mathcal{M}}$ as:\cite{Pompeo_Part1}
\begin{equation}
\label{eq:MB}
\dbar{\mathcal{M}}(\theta,\phi)=-\xma{B}(\theta,\phi)\dbar{{M}}^{-1}\xma{B}(\theta,\phi)
\end{equation}
and where:
\begin{equation}
\label{eq:epsilon}
\epsilon^2(\theta)=\cos^2\theta+\gamma^{-2}\sin^2\theta
\end{equation}
is the square of the well-known angular-dependent anisotropy parameter\cite{blatterone, tinkham}  which defines, among the others, the angle dependence of the critical field $B_{c2}(\theta)=B_{c2}(0)/\epsilon(\theta)$.
From Eq. \eqref{eq:rho_v} it can be noted that, contrary to the intrinsic resistivity tensor $\dbar{\rho}_{v}^\i$, the measurable tensor $\dbar{\rho}_{v}$ does not obey the angular scaling law, since it incorporates additional angular dependencies through $\dbar{\mathcal{M}}$ and $\epsilon^2$.

Now I include the effects of flux creep. The pinning energy $U$ depends only on the magnetic field magnitude and direction and not on the direction of vortex motion. Moreover, it obeys the usual scaling law (with a constant scaling factor\cite{BGL} $\gamma^{-1}$). Therefore the thermal depinning time $\tau_{th}$ (Eq. \eqref{eq:tauth}) is a scalar $\omega_p=\tau_p^{-1}$. The pinning constant intrinsic tensor (Eq. \eqref{eq:kp1}) can thus be corrected to include creep as:
\begin{equation}
\label{eq:kpicreep}
\dbar{k}_{p}^\i\!(B,\theta)\!=\!k_{p,11}^\i(B/B_{c2}(\theta))\left(\!1\!-\frac{\rmi}{\omega\tau_{th}\!(B/B_{c2}(\theta))}\right)\!\dbar{M}^{-1}
\end{equation}
Consequently, the scalar vortex motion resistivity within flux creep, Eq. \eqref{eq:rhoBiso}, can be generalized to the anisotropic case as follows, yielding an expression analogous to Eq. \eqref{eq:rho_vi2}:
\begin{equation}
\label{eq:rho_vi3}
\dbar{\rho}_v^\i(B,\theta)=\rho_\ff^\i\frac{\varepsilon'+\rmi\frac{\omega}{\omega_{0}}}{1+\rmi\frac{\omega}{\omega_{0}}}\dbar{M}=\rho_{v,11}^\i(B/B_{c2}(\theta))\dbar{M} 
\end{equation}
where the characteristic frequency $\omega_0$ and the creep factor $\varepsilon'$ remain scalar values as in the isotropic case, reported in Eqs. \eqref{eq:omega0} and \eqref{eq:creep}.
This property will prove important in the interpretation of the experiments.

Finally, the remaining quantity to be derived is the complex mobility tensor $\dbar{\mu}_v$, which is of particular interest for its application within the full electrodynamic model to be treated in Section \ref{sec:CCmain}. Since the quantity which appears in the electrodynamics expressions is actually $(\xma{B}\dbar{\mu}_v\xma{B})$,\cite{Pompeo_Part1} it is sufficient to consider the following implicit expression of $\dbar{\mu_v}$ in terms of $\dbar{\rho_v}$:\cite{Pompeo_Part1}
\begin{equation}
\label{eq:mu_complex}
\dbar{\rho}_v=-\xma{B}\left(\dbar{\mu}_v\Phi_0B\right)\xma{B}
\end{equation}
Incidentally, it is interesting to observe that by comparing Eq. \eqref{eq:mu_complex} to Eq. \eqref{eq:rho_v}, one has: 
\begin{equation}
\label{eq:rhoffequivl}
\xma{B}\left(\dbar{\mu}_v\Phi_0B\right)\xma{B}=\xma{B}\left(\frac{\rho_{v,11}^\i}{\epsilon^2}\dbar{M}^{-1} \right)\xma{B}
\end{equation}
The above identity does not allow to equate the tensor between round parentheses because the matrix $\xma{B}$ is not invertible.   
Actually it can be easily verified that $(\dbar{\mu}_v\Phi_0 B)\neq(\rho_{\ff,11}^\i/\epsilon^2\dbar{M}^{-1})$. Nevertheless, Eq. \eqref{eq:rhoffequivl} shows that they are physically equivalent when they are involved in the electrodynamics expressions, which always include the ``sandwich'' product with $\xma{B}$. Indeed, this fact can be visually shown by expressing Eq. \eqref{eq:rhoffequivl} in a rotated frame of reference\cite{Pompeo_Part1} where the $z$-axis coincides with the magnetic field direction $\hat{B}$. In this frame, the two matrices under consideration become identical apart from the elements belonging to the third row and to the third column. Considering that these elements do not affect the ``sandwich'' product with the rotated $\xma{B}$, the equivalence from a physical electrodynamics point of view of the two matrices is understood.

Before concluding this Subsection, it is worth to stress that the choice of a relaxation of the pinning constant\cite{BrandtModel} to deal with flux creep is not a limiting factor to the results obtained up to now: in fact, any pure thermal creep process (independent on the angle between the field and the current), possibly with a different definition of $\varepsilon'$ and $\omega_0$ (Ref. \onlinecite{PompeoPRB}), would yield the same results.

\section{Application to experiments: the measured complex vortex resistivity in common setups}
\label{sec:rhovm_examples}

In this Section I consider explicitly some typical experimental configurations and I derive specific expressions relating measured and intrinsic properties.

\subsubsection{Straight planar currents}
\label{sec:rhovm_examples1}
The resistivity measured along the current direction $\hat{J}$ can be computed as:\cite{Pompeo_Part1}
\begin{equation}
\label{eq:rhoJ}
\rho^{(\widehat{J})}=\left(\dbar{\rho}\widehat{J}\right)\cdot\widehat{J}
\end{equation}
By applying a straight a.c. current $\vec{J}\parallel\hat{x}$ and using the resistivity tensor given by Eq. \eqref{eq:rho_v}, one obtains for the measured vortex resistivity:
\begin{subequations}
\label{eq:rhoeff_exp4}
\begin{align}
\rho_{v}^{(x)}(B,\theta,\phi)&=\rho_{v,11}^\i(B/B_{c2}(\theta))f_{L\phi}(\theta,\phi)\\
f_{L\phi}(\theta,\phi)&=\frac{\gamma^{-2}\sin^2\theta\sin^2\phi+\cos^2\theta}{\gamma^{-2}\sin^2\theta+\cos^2\theta} 
\end{align}
\end{subequations}
I recall that $\phi$ is the angle between the projection of the $\vec{B}$ field on the $x$-$y$ plane and the $x$ axis (Fig. \ref{fig:ref}). It can be seen that the effective, measured vortex motion resistivity consists in the product of two terms: the first one relates to the intrinsic resistivity $\rho_{v,11}^\i$ only, which in particular obeys the angular scaling law; the second one, denoted in the equation as $f_{L\phi}(\theta,\phi)$, contains an ``extrinsic'' angular dependence which arises from the Faraday and Lorentz actions. Consequently, as already anticipated commenting the whole tensor $\dbar{\rho}_v$, the experimentally measured quantity does not obey the scaling law.
Only in the case $\phi=\pi/2$  $(\vec{B}\perp\vec{J})$ one has direct, experimental access to the intrinsic vortex resistivity, which obeys the angular scaling law.

\subsubsection{Rotational symmetric planar currents}
\label{sec:rhovm_examples2}
\begin{figure}[ht]
\centerline{\includegraphics[width=8.5cm]{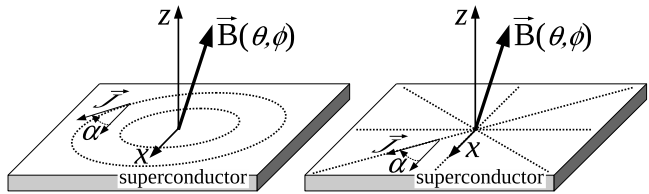}}
  \caption{ Circular (left panel) and radial (right panel) symmetric current patterns. $\alpha$ is the angle between the (local) current density $\vec{J}$ and the $x$ axis.}
\label{fig:rotcurrents}
\end{figure}

A rotational symmetric planar geometry is often used for measurements of the vortex-state microwave response. Examples are cylindrical resonators,\cite{tsuchiya,cylres} in which (using cavity perturbation techniques\cite{perturbation}) the superconducting sample is located on the circular bases, and the so-called Corbino disk setup,\cite{creepCorbino,Corbino} in which the superconducting sample short-circuits an open-ended coaxial cable. In both cases, rotational symmetric currents (circular and radial for the resonator and the Corbino disk, respectively) are induced along the superconductor $a$-$b$ plane (see Fig. \ref{fig:rotcurrents}), in the frequent case in which the superconductor $c$-axis is perpendicular to its surface.
The effective measured\cite{circurrent,circurrent2} resistivity (see the following section for the relationship between measured surface impedance and the resistivity) comes out as an angular average\cite{Collin} over the current pattern:
\begin{align}
\label{eq:rhoeff_exp51}
\nonumber
\rho_{v}^{(\circ)}&(B,\theta)=\frac{1}{2\pi}\int_0^{2\pi}\left(\dbar{\rho}_v(B,\theta,\phi)\widehat{J}(\alpha)\right)\cdot\widehat{J}(\alpha) d\alpha=\\
=&\frac{1}{2\pi}\frac{\rho_{v,11}(B,\theta)}{\epsilon^2(\theta)}\int_0^{2\pi}\left(\dbar{\mathcal{M}}(\theta,\phi)\widehat{J}(\alpha)\right)\cdot\widehat{J}(\alpha) d\alpha
\end{align}
where $\alpha$ is the angle between the (local) current density $\vec{J}$ and the $x$ axis (see Fig. \ref{fig:rotcurrents}).
Exploiting the uniaxial anisotropy together with the circular symmetry of the current pattern, the computation can be equivalently and more simply done by averaging Eq. \eqref{eq:rhoeff_exp4} over all possible values of $\phi$:
\begin{align}
\nonumber
\rho_{v}^{(\circ)}(B,\theta)&=\frac{1}{2\pi}\int_0^{2\pi}\rho_{v}^{(x)}(\theta,\phi) d\phi
\end{align}
The result is:
\begin{subequations}
\label{eq:rhoeff_exp53}
\begin{align}
\label{eq:rhoeff_exp53a}
\rho_{v}^{(\circ)}(B,\theta)&=\rho_{v,11}^\i(B/B_{c2}(\theta))f_{L}(\theta)\\
f_{L}(\theta)&=\frac{\frac{1}{2}\gamma^{-2}\sin^2\theta+\cos^2\theta}{\gamma^{-2}\sin^2\theta+\cos^2\theta}
\end{align}
\end{subequations}
It can be noted that there is no more the angular dependence on $\phi$, and that the function $f_{L}(\theta)$ is a particularization of the above introduced $f_{L\phi}$. Additionally, the same comments proposed for Eq. \eqref{eq:rhoeff_exp4} hold also here. Moreover, one can note that, in the present case, if $\rho_{v,11}^\i\propto(B/B_{c2}(\theta))^\beta$, Eq. \eqref{eq:rhoeff_exp53} still yields a scaling law.\cite{circurrent2,submittedPompeo} However, in the interpretation of the experiments one has to be careful and not confuse this artificial scaling function with the theoretical scaling expression.

\subsubsection{Angle-dependent effective quantities}
It is interesting to note that a typical GR model analysis in a isotropic superconductor extracts the vortex parameters $\rho_\ff$, $r$ and $k_p$ from the complex measured $\rho_{v}$ of Eq. \eqref{eq:rhoGRiso} as follows:
\begin{align*}
\label{eq:rhoGR2}
  \rho_{\ff} & =\Re(\rho_{v}) \left[1+\left(\frac{\Im(\rho_{v})}{\Re(\rho_{v})}\right)^2\right] \\
  r &=\frac{\Im(\rho_{v})}{\Re(\rho_{v})} \\
  k_p &=\frac{r}{\rho_\ff}\omega B \Phi_0 = \omega B \Phi_0 \frac{\Im(\rho_{v})}{\Re^2(\rho_{v})+\Im^2(\rho_{v})}
\end{align*}
On the other hand, when performing measurements on an anisotropic superconductor probed with the circular current pattern leading to Eq. \eqref{eq:rhoeff_exp53} (apart from the additional $\phi$-dependence, the same holds obviously for the straight current setup related to Eq. \eqref{eq:rhoeff_exp4}), this computation would yield the following {\em effective} quantities:
\begin{subequations}
\label{eq:rhoGR3}
\begin{align}
  \rho_{\ff,\eff}(B,\theta) & =\rho_{\ff,11}^\i(B/B_{c2}(\theta)) f_{L}(\theta) \\
  r_{\eff}(B,\theta) &=r(B/B_{c2}(\theta)) \\
  k_{p,\eff}(B,\theta) &=\frac{k_{p,11}^\i(B/B_{c2}(\theta))}{f_{L}(\theta)}
\end{align}
\end{subequations}
It can be seen that the parameter $r=\omega_p/\omega$ is directly obtained from the measured quantities: this is an interesting result, which allows a direct evaluation of the material anisotropy of the system without the need to deal with Lorentz-dependent contribution $f_{L}(\theta)$. On the other hand, both $\rho_{\ff,\eff}$ and $k_{p,\eff}$ show an additional angular dependence through $f_{L}(\theta)$. Therefore, in the analysis of angular data care must be devoted in correctly extracting the intrinsic quantities instead of the effective ones: this requires to evaluate in some way the $f_{L}(\theta)$ function, which in turn requires the knowledge of the anisotropy factor $\gamma$.

Further comments can be done considering a scaling analysis performed starting from the effective quantities of Eqs. \eqref{eq:rhoGR3}. 
Once the intrinsic quantities $\rho_{\ff,11}^\i$ and $r$ are extracted, they can be checked against the scaling prescription. If $\rho_{\ff,11}^\i$ is found to satisfy the scaling, and in the same time $r$ or, equivalently $\rho_{C,11}^\i$, is not, this result would unambiguously indicate the presence of directional pinning contributions such as extended defects.

\section{High frequency electrodynamic response in anisotropic superconductors}
\label{sec:CCmain}

Having discussed the angular dependent vortex motion resistivity, I now include the effect of superfluid and quasiparticles in the overall electromagnetic response. 
In the high frequency regime, an electromagnetic wave impinging on a superconductor, driven in the mixed state by a static magnetic field $\vec{B}$, determines an electromagnetic response which is not only dictated by the motion of the vortices but arises from the coupling between the latter and the high frequency currents, which include both the normal and the superconducting components.\cite{CCiso,PlacaisPRB54} 

Moreover, the actual physical quantity that can be directly measured is not the mixed state complex resistivity, but instead the superconductor surface impedance\cite{rparameter1,tsuchiya,rparameter2,examplesZthin,examplesZ} $Z$, defined as the ratio between the components of the electric and magnetic fields tangential to the separation surface between the superconductor and the outer medium (typically vacuum or air). While the complex resistivity, in the local response regime here considered, can be computed as the local relation between the electric field and the current density, the evaluation of the surface impedance $Z$ typically requires the determination of the electromagnetic field distribution within the whole superconductor. In tilted static magnetic fields the computation is far from trivial. Coffey and Clem (CC)\cite{CCiso,CCisoblique, CC11757, CCaniso, coffeyJLTP} used an isotropic description of vortex dynamics, with scalar vortex mobility, to calculate the response of isotropic superconductors 
in 
parallel \cite{CCiso} and oblique static magnetic fields,\cite{CCisoblique,CC11757} as well as of anisotropic superconductors in  arbitrary oriented magnetic fields.\cite{CCaniso,coffeyJLTP}
In the following, (i) I calculate the anisotropic, angular dependent response to electromagnetic fields in the mixed state by including the full tensor model for the vortex dynamics developed in the previous Section (Section \ref{sec:CCmodel}) and (ii) I provide an example of application by computing the surface impedance tensor (whose definition is recalled in Section \ref{sec:Zaniso}) of an anisotropic superconductor in the commonly found thin film geometry\cite{examplesZthin,rparameter2} (Section \ref{sec:Zfilmaniso}).

\subsection{The fully coupled electrodynamics anisotropic model}
\label{sec:CCmodel}
The coupling of vortex motion, superfluid and quasiparticles has been addressed, among the others,\cite{CCiso, PlacaisPRB54} by Coffey and Clem.
The core of the CC model is a partial differential equation whose solution yields the high-frequency part $\vec{b}$ of the magnetic field induction existing within the superconductor. Within the linear response regime, for an anisotropic superconductor in the $e^{\rmi\omega t}$ sinusoidal regime, under the effect of a static and uniform magnetic field $\vec{B}$ with rigid vortices, the equation in the tensor approach here used reads:
\begin{align}
\label{eq:PDE1}
\nonumber
-\nabla\times\left(\dbar{\lambda}^2\left(\nabla\times\vec{b}\right)\right)+\mu_0\nabla\times\left(\dbar{\lambda}^2\dbar{\sigma}_{\nf}\vec{E}\right)-\vec{b}=\\
-\rmi\frac{\Phi_0B}{\omega\mu_0} \nabla\times\left(\hat{B}\times\left(\dbar{\mu}_v\left(\left(\nabla\times\vec{b}\right)\times\hat{B}\right)\right)\right)
\end{align}
The uniform static field $\vec{B}$ and the high frequency $\vec{b}$ yield the total magnetic induction $\vec{B}_{tot}=\vec{B}+\vec{b}$.

The tensors $\dbar{\lambda}$ and $\dbar{\sigma}_{\nf}$ are the anisotropic London penetration depth and the normal fluid conductivity, respectively. 
By expressing them in terms of the mass anisotropy only, thus neglecting any scattering time anisotropy and taking the Cooper pair mass anisotropy equal to the electronic mass anisotropy\cite{Bespalov} $\dbar{M}$, one has: 
\begin{subequations}
\begin{align}
\label{eq:lambda_sigmanf}
\dbar\lambda^2&=\lambda^2_{11}\dbar{M} \\
\dbar\sigma_{\nf}&=\sigma_{\nf,11}\dbar{M}^{-1} 
\end{align}
\end{subequations}
where, with the same notation of Eq. \eqref{eq:masstensor}, $\lambda^2_{11}$ and $\sigma_{\nf,11}$ denote respectively the squared London penetration depth and the normal-fluid conductivity in the $a$-$b$ plane. 
Then, with Eqs. \eqref{eq:rhoffequivl} and \eqref{eq:rho_v} for $\dbar{\mu}_v$ and $\dbar{\rho}_v$, Eq. \eqref{eq:PDE1} becomes: 
\begin{align}
\label{eq:PDE}
\nonumber
\beta\vec{b}-\lambda^2_{11} \nabla\times\left(\dbar{M}\left(\nabla\times\vec{b}\right)\right)=\\
-\frac{\rmi}{2}\delta_{v,11}^2 \nabla\times\left(\hat{B}\times\left(\frac{\dbar{M}^{-1}}{\epsilon^2}\left(\left(\nabla\times\vec{b}\right)\times\hat{B}\right)\right)\right)
\end{align}
where 
\begin{equation}
\label{eq:PDEbeta}
\beta=(-{2\rmi\lambda^2_{11}}/{\delta_{nf,11}^2}-1)
\end{equation}
and
the quantities $\delta_{nf,11}$ and $\delta_{v,11}$ are respectively the normal fluid and vortex motion penetration depths in the $a$-$b$ plane:
\begin{subequations}
\label{eq:depths}
\begin{align}
\delta_{nf,11}^2&=\frac{2}{\omega\mu_0\sigma_{nf,11}} \\
\delta_{v,11}^2&=\frac{2\rho_{v,11}}{\omega\mu_0}
\end{align}
\end{subequations}
Equation \eqref{eq:PDE}, by including the tensor description of the vortex dynamics  $\dbar{\mu}_v$ and $\dbar{\rho}_v$ developed in the previous Section, is a very general formulation of the mixed state electromagnetic response, and it includes the effects of the anisotropy and of the angle between $\vec{B}$ and $\vec{J}$ (see Eqs. \eqref{eq:rho_v} and \eqref{eq:mu_complex}).

The high frequency field $\vec{b}$ can be completely determined through Eq. \eqref{eq:PDE} with the appropriate boundary conditions.\cite{CC11757,coffeyJLTP} Other quantities of interest are the total (superconducting plus normal fluid) current density, which can be derived by using the fourth Maxwell equation $\nabla\times\vec{b}=\mu_0 \vec{J}$ (displacement current is neglected as ordinarily done in superconductors \cite{coffeyJLTP}), and the high-frequency electric field $\vec{E}$, which can be derived from the London equation yielding:
\begin{equation}
\label{eq:rhoCC_def}
\vec{E}=\dbar{\rho}_{cc}\vec{J}
\end{equation}
where $\dbar{\rho}_{cc}$ is the Coffey\textendash{}Clem-like resistivity tensor:\cite{coffeyJLTP}
\begin{align}
\label{eq:rhocc} 
\dbar{\rho}_{cc}
&=\frac{\rmi\omega\mu_0\lambda^2_{11}\dbar{M}+\rho_{v,11}^\i\frac{1}{\epsilon^2}(-\xma{B}\dbar{M}^{-1}\xma{B})}{1+\frac{2\rmi\lambda^2_{11}}{\delta_{nf,11}^2}}
\end{align}
where the vortex resistivity in the form of the tensor $\dbar{\rho}_v$ has been explicitly included.

A significant limit of the above expression is obtained for temperatures $T \rightarrow 0$: in this case $\sigma_{nf,11}\rightarrow 0$, $\delta_{nf,11}\rightarrow \infty$, so that:
\begin{equation}
\label{eq:rhocclimit} 
\dbar{\rho}_{cc}\rightarrow \rmi\omega\mu_0\lambda^2_{11}\dbar{M}+\frac{\rho_{v,11}^\i}{\epsilon^2}(-\xma{B}\dbar{M}^{-1}\xma{B})
\end{equation}
which shows that at low temperatures the whole complex resistivity is dominated by vortex motion and supercurrents and which is the anisotropic variant of the isotropic Brandt's full model.\cite{BrandtModel}
 
Once both $\vec{b}$ and $\vec{E}$ are known, the surface impedance tensor $\dbar{Z}$ can be computed following the definition valid for anisotropic media recalled in the next Section.

\subsection{Surface impedance for anisotropic media}
\label{sec:Zaniso}
For an anisotropic media the surface impedance $\dbar{Z}$ is a two-dimensional tensor defined through the following expression:\cite{Zaniso}
\begin{equation}
\label{eq:Zaniso} 
\vec{E}_{\parallelslant}=\dbar{Z}\left(\hat{n}\times \vec{h}_{\parallelslant}\right)
\end{equation}
where $\hat{n}$ is the normal surface enclosing the media and pointing outwards, toward the surrounding vacuum or air, $\vec{h}$ is the magnetic field equal to $\vec{b}/\mu_0$ since $\mu_r=1$, and the subscript ``$\parallelslant$'' denotes the vector components parallel to the surface. Placing the superconductor in the positive z-semi-space, its delimiting surface is at $z=0$ and in the above equation one has $\hat{n}=-\hat{z}$, $\vec{E}_{\parallelslant}=(E_x, E_y)$ and $\vec{h}_{\parallelslant}=(h_x, h_y)$.
Using the fourth Maxwell equation $\nabla\times\vec{h}= \vec{J}$ along with $\vec{E}=\dbar{\rho}\vec{J}$, one writes:
\begin{equation}
\label{eq:EfromH} 
\vec{E}=\dbar{\rho}\nabla\times\vec{h}
\end{equation}
Combining Eqs. \eqref{eq:Zaniso} and \eqref{eq:EfromH}, one obtains:
\begin{equation}
\label{eq:Zfromrho} 
\dbar{Z}=
\begin{pmatrix}
\rho_{11} & \rho_{12} \\
\rho_{21} & \rho_{22} \\
\end{pmatrix}
\begin{pmatrix}
-\frac{1}{b_y}\frac{\partial b_y}{\partial z} & \frac{1}{b_x}\frac{\partial b_y}{\partial z} \\
\frac{1}{b_y}\frac{\partial b_x}{\partial z} & -\frac{1}{b_x}\frac{\partial b_x}{\partial z} \\
\end{pmatrix}_{\!\!\!z=0}
\end{equation}
From this equation it is clear that in order to compute the surface impedance tensor both the superconductor resistivity tensor and the electromagnetic field distribution in the superconductor are needed. 

\subsection{Application to experiments: surface impedance for superconducting anisotropic thin film in titled field}
\label{sec:Zfilmaniso}
In this section the expression for the surface impedance of an anisotropic superconductor in tilted magnetic field within the CC model, including the anisotropic vortex dynamics description developed in Section \ref{sec:rhovmtensor}, will be derived for a thin film geometry, which is  encountered in a large number of experimental studies.\cite{examplesZthin,rparameter2} 

\begin{figure}[ht]
\centerline{\includegraphics[width=4cm]{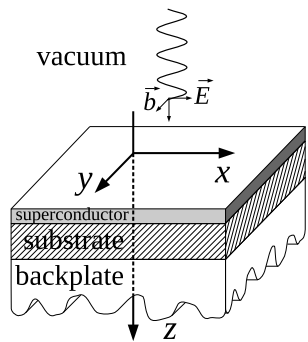}}
  \caption{ Thin film, grown on dielectric substrate, with backing metal plate.} 
\label{fig:thinfilm}
\end{figure}

Superconducting thin films grown on dielectric substrates are partially transparent to the electromagnetic radiation, so that in microwave measurements they are customarily backed with a metallic plate. The resulting layered structure is depicted in Fig. \ref{fig:thinfilm}. The structure is considered to be indefinitely extended along the $x$-$y$ directions, thus neglecting border effects.

Considering an impinging TEM wave propagating on the z direction, the high frequency fields are uniform on the $x$-$y$ planes. The surface impedance tensor $\dbar{Z}$ can be computed by applying the model presented in Sec. \ref{sec:CCmodel}. The full computation is provided in the Appendix \ref{sec:CCaniso_imp}. 

Here I report the result: within the well-known thin film approximation \cite{thinfilm,thinfilm2} $t_s\ll\min(\delta_{nf}, \lambda)$ (being $t_s$ the film thickness), one has:
\begin{equation}
\label{eq:Zthin_aniso_within_text} 
\dbar{Z}=\frac{1}{t_s}
\begin{pmatrix}
\rho_{cc,11} & \rho_{cc,12} \\
\rho_{cc,21} & \rho_{cc,22} \\
\end{pmatrix}
\end{equation}
which appears as an intuitive (but not straightforward) generalization of the corresponding isotropic expression $Z=\rho_{cc}/t_s$. 

For the thin film geometry here considered, it can be noted that the surface impedance tensor $\dbar{Z}$ is directly proportional to the resistivity (sub)tensor. In particular, this direct proportionality allows for the immediate generalization to $\dbar{Z}$ of the results concerning the measurable resistivities discussed in Section \ref{sec:rhovm_examples}, as well as a direct separation of the real and imaginary parts of $\dbar{\rho}$ in the real and imaginary parts of $\dbar{Z}$.
On the other hand, for thicker films the expression becomes quickly more cumbersome: by examining the next higher order terms reported in Eq. \eqref{eq:z_tensor_elements}, it can be seen that the $\dbar{Z}$ elements do not correspond to single elements of $\dbar{\rho}_{cc}$. Instead the elements of $\dbar{Z}$ become a function of the elements of $\dbar{\rho}_{cc}$m through complex coefficients, making more difficult the extraction of the resistivity from the measured surface impedance.

\section{Summary}
In this paper a full tensor model for the a.c. vortex motion resistivity, including creep, pinning and flux flow, has been presented. The model has been employed to develop a full tensor theory for the electromagnetic response of an uniaxially anisotropic type-II superconductors in the vortex state. Arbitrary orientations between the applied field, the applied current and the anisotropy axis have been considered. Limiting cases have been commented. Relations between measurable quantities and intrinsic material properties have been given, showing that care must be put in separating the material angular dependence from the one arising from the geometry of the setup. Finally, the expression of the measured surface impedance in the largely used thin film geometry has been computed.

\acknowledgments
This work has been supported by Regione Lazio and partially supported by the Italian FIRB project ``SURE:ARTYST'' and by EURATOM.

The author warmly acknowledges the helpful discussion with prof. E. Silva.

\appendix
\section{Application to anisotropic thin films}
\label{sec:CCaniso}
In this Appendix the computation of the surface impedance tensor of the thin film geometry described in Sec. \ref{sec:Zfilmaniso} is fully developed, along the lines of Refs. \onlinecite{CCaniso,CC11757,coffeyJLTP}, where similar computations for different geometries were performed. For the ease of comparison, whenever possible the same notation will be used. Moreover, in the following the subscript ``11'' for the $a$-$b$ plane quantities $\lambda_{11}$ and $\delta_{v,11}$ is dropped in order to simplify the notation.

\subsection{Geometry of the problem}
\label{sec:CCaniso_geom}
The geometry here considered is represented in Fig. \ref{fig:thinfilm}, with the $z$ axis normal to the superconductor surface $z=0$ and pointing inward.
The impinging TEM electromagnetic field is assumed to determine a field:
\begin{equation}
\label{eq:bhfboundary} 
\vec{b}(z=0)=b_0 \hat y
\end{equation}
(with $b_0\ll B_0$) on the superconductor surface, thus defining the first boundary condition.

Within the superconductor of thickness $t_s$, i.e. for $0\leq z \leq t_s$, the field will be in general:
\begin{equation}
\label{eq:bhf} 
\vec{b}(z)=f_1(z) \hat x+f_2(z) \hat y
\end{equation}
Concerning the second boundary, typically the (effective) surface impedance of the film substrate is much larger than the characteristic surface impedance of the superconductor,\cite{thinfilm,thinfilm2} a part from very specific cases in which the substrate give rise to more complex phenomena.\cite{thinfilm2,substrateeffects} At the interface between the superconductor and the substrate, i.e. at $z=t_s$, the reflection coefficient for the electric field is $\simeq1$. Hence the tangential component of $\vec{b}$ is $\simeq0$ at the interface.\cite{Johnk} 

Hence, the overall boundary conditions are the following:

\begin{subequations}
\label{eq:PDEbound} 
\begin{eqnarray}
f_1(0)=0&,&f_2(0)=b_0 \\
\label{eq:PDEbound_b} 
f_1(t_s)=0&,&f_2(t_s)=0
\end{eqnarray}
\end{subequations}

\subsection{High frequency magnetic field computation}
\label{sec:CCaniso_PDE}
Applying Eq. \eqref{eq:PDE} to this configuration and separating the two components of $\vec{b}$, one has:
\begin{subequations}
\label{eq:PDEsep} 
\begin{eqnarray}
a f_1^{''}(z)+\beta f_1(z)+c f_2^{''}(z)&=&0 \\
{d} {f}_2^{''}(z)+\beta f_2(z)+c f_1^{''}(z)&=&0
\end{eqnarray}
\end{subequations}
where $\beta$ is defined in Eq. \eqref{eq:PDEbeta} and the parameters $a$, $d$ and $c$ are defined as:
\begin{subequations}
\label{eq:PDEpara} 
\begin{eqnarray}
a&=&\lambda^2-\frac{\rmi}{2}\delta_v^2(\gamma^{-2}B_{x}^2+B_{z}^2)\frac{1}{\epsilon^2} \\
d&=&\lambda^2-\frac{\rmi}{2}\delta_v^2(\gamma^{-2}B_{y}^2+B_{z}^2)\frac{1}{\epsilon^2} \\
c&=&-\frac{\rmi}{2}\delta_v^2 \gamma^{-2}B_{x} B_{y}\frac{1}{\epsilon^2}
\end{eqnarray}
\end{subequations}
As it can be seen, the parameters $a$, $c$ and $d$ depend on the superconductor properties and on the static magnetic field orientation, here written in compact way as $\hat{B}=(B_{x}, B_{y}, B_{z})$.
Equations \eqref{eq:PDEsep} and Eqs. \eqref{eq:PDEpara}, written for the thin film geometry, are similar to those reported in Refs. \onlinecite{CCaniso,coffeyJLTP}. In Eqs. \eqref{eq:PDEsep} and \eqref{eq:PDEpara} the vortex mobility tensor has been used.

The coupled differential Eqs. \eqref{eq:PDEsep}, with the boundary conditions, Eqs. \eqref{eq:PDEbound}, can be solved by means of the Laplace transformation $F_i(s)=\int_0^\infty f_i(z)e^{-s z}dz$ and anti-transforming the solutions using the partial fraction expansion. 
One obtains $f_1(z)$ and $f_2(z)$ as linear combinations of the four exponentials $e^{\pm z/\lambda_\pm}$ with coefficients $A_{j,k}$ (with $j=1..4$ and $k=1,2$) which arise from the fraction expansion:
\begin{align}
\label{eq:f_fullexp}
\nonumber
f_k(z)&=A_{1,k}e^{z/\lambda_-}+A_{2,k}e^{-z/\lambda_-}+\\
&+A_{3,k}e^{z/\lambda_+}+A_{4,k}e^{-z/\lambda_+}
\end{align}
The expressions of the coefficients $A_{j,k}$, not reported here, include the first derivatives $f_1'(0)$ and $f_2'(0)$ arising from the Laplace transformation. 
The pair of complex penetration depths  $\lambda_\pm$ is given by the solutions $s^2=-\lambda^{-2}_\pm$ of the quadratic equation $(ad-c^2)s^4+\beta(a+d)s^2+\beta^2=0$. Their explicit expressions are:
\begin{subequations}
\label{eq:lambdapm}
\begin{align}
\lambda_{+}^{-2}&=\frac{-\beta}{\lambda^2-\frac{\rmi}{2}\delta_v^2} \\
\lambda_{-}^{-2}&=\frac{-\beta}{\lambda^2-\frac{\rmi}{2}\delta_v^2\frac{B_{z}^2}{\epsilon^2}}
\end{align}
\end{subequations}
In the well known\cite{thinfilm,thinfilm2} thin film approximation, $t_s\ll\min(\delta_{nf}, \lambda)$, one easily obtains $|t_s/\lambda_{\pm}|\ll|\sqrt{\beta}t_s/\lambda|\ll1$. Within this condition, the exponentials in Eq. \eqref{eq:f_fullexp} can be Taylor-expanded and truncated to the first terms, allowing also to easily compute $f_1'(0)$ and $f_2'(0)$ through the application of the boundary condition \eqref{eq:PDEbound_b}:
\begin{subequations}
\label{eq:f_taylor}
\begin{align}
\!\!f_1(z)&\cong b_0 \frac{c\beta t_s^2}{2(ad-c^2)}\left(-\frac{z}{t_s}+\frac{z^2}{t_s^2}\right) \\ 
\!\!f_2(z)\!&\cong b_0 \left[1+\frac{z}{t_s}\left( \frac{a\beta t_s^2}{2(ad-c^2)}-1\!\right)\!-\frac{z^2}{t_s^2}\frac{a\beta t_s^2}{2(ad-c^2)}\right]
\end{align}
\end{subequations}
Truncation to the second order term avoids $f_1(z)=0$, a too crude approximation.

The explicit values of the derivatives at $z=0$, needed for the computation of the surface impedance according to the definition \eqref{eq:Zfromrho}, are reported here:
\begin{subequations}
\label{eq:f_der}
\begin{align}
f_1'(0)&=-b_0\frac{c\beta t_s}{2(ad-c^2)} \\
f_2'(0)&=-b_0\left(\frac{1}{t_s}-\frac{a\beta t_s}{2(ad-c^2)}\right)
\end{align}
\end{subequations}

\subsection{Surface impedance computation}
\label{sec:CCaniso_imp}
It is now possible to explicitly compute the surface impedance elements, applying the definition Eq. \eqref{eq:Zfromrho} with the resistivity tensor given by Eq. \eqref{eq:rhocc} and with the results of Eqs. \eqref{eq:f_taylor} and \eqref{eq:f_der}:
\begin{subequations}
\label{eq:z_tensor_elements}
\begin{align}
Z_{11}=& \frac{\rho_{cc,11}}{t_s}-\frac{\beta t_s(a\rho_{cc,11}+c\rho_{cc,12})}{2(ad-c^2)}\cong\frac{\rho_{cc,11}}{t_s} \\
Z_{21}=& \frac{\rho_{cc,21}}{t_s}-\frac{\beta t_s(a\rho_{cc,21}+c\rho_{cc,22})}{2(ad-c^2)}\cong\frac{\rho_{cc,21}}{t_s}
\end{align}
\end{subequations}
where in the last approximate equalities the higher order terms of the Taylor expansion have been finally neglected. 

Now, by solving again the differential equations with $\vec{b}(z=0)=b_0 \hat{x}$, one obtains the expressions equivalent to \eqref{eq:z_tensor_elements} for the elements $Z_{12}$ and $Z_{22}$, so that it is finally demonstrated that:
\begin{equation}
\label{eq:Zthin_aniso} 
\dbar{Z}=\frac{1}{t_s}
\begin{pmatrix}
\rho_{cc,11} & \rho_{cc,12} \\
\rho_{cc,21} & \rho_{cc,22} \\
\end{pmatrix}
\end{equation}


\begin{thebibliography}{99}
\label{sec:TeXbooks}%

\bibitem{anisotropicSC}
P. M. Aswathy, J. B. Anooja, P. M. Sarun, and U. Syamaprasad, \SUST {\bf 23}, 073001 (2010); 
J. Paglione and R. L. Greene, \NatPhys {\bf 6}, 645 (2010); H. Nakamura, M. Machida, T. Koyama, and N. Hamada, \JPSJ{\bf 78} 123712.1 (2009); 
C. Buzea and T. Yamashita, \SUST {\bf 14}, R115 (2001);
C. P. Poole Jr., H. A. Farach, R. J. Creswick, and R. Prozorov, {\it Superconductivity, 2nd Ed.} (Academic Press, Netherlands, 2007).

\bibitem{Pompeo_Part1} N. Pompeo, {\it Theory of measurements of electrodynamic properties in anisotropic superconductors in tilted magnetic fields. Part I: flux flow and Campbell regimes}, submitted for publication (2012), arXiv:1211.3355 [cond-mat.supr-con].

\bibitem{previousWorks} B. I. Ivlev and N. B. Kopnin, \EPL {\bf 15}, 349 (1991); V. M. Genkin and A. S. Mel'nikov, \JETP {\bf 68}, 1254 (1989); Z. Hao, C-Ren Hu, and C.-S. Ting, \PRB {\bf 51}, 9387 (1995); Z. Hao, C-Ren Hu, and C.-S. Ting, \PRB {\bf 52}, R13138 (1995); Z. Hao, C-Ren Hu, and C.-S. Ting, \PRB {\bf 48}, 16818 (1993); Z. Hao and C-Ren Hu, \JLTP {\bf 104}, 265 (1996); E. H. Brandt, \ZPB {\bf 80}, 167 (1990); Th. Klupsch, \PC {\bf 197}, 224 (1992); V. A. Shklovskij  and O. V. Dobrovolskiy, \PRB {\bf 74}, 104511 (2006).

\bibitem{haoclem} Z. Hao and J. R. Clem, \IEEEmag {\bf 27}, 1086 (1991).

\bibitem{CCaniso} M. W. Coffey, \PRB {\bf 47}, 12284 (1993).

\bibitem{coffeyJLTP} M. W. Coffey, \JLTP {\bf 108}, 331 (1997). 

\bibitem{BrandtModel} E. H. Brandt, \PRL {\bf 67}, 2219 (1991).

\bibitem{CCiso} M. W. Coffey and J. R. Clem, \PRL {\bf 67}, 386 (1991); \PRB 45, {\bf 9872} (1992).

\bibitem{creepCorbino} E. Silva, N. Pompeo and S. Sarti, \SUST {bf 24}, 024018 (2011).

\bibitem{APL91} N. Pompeo, R. Rogai, E. Silva, A. Augieri, V. Galluzzi and G. Celentano, \APL {\bf 91}, 182507 (2007).

\bibitem{SongPRB79} C. Song, T. W. Heitmann, M. P. DeFeo, K. Yu, R. McDermott, M. Neeley, John M. Martinis and B. L. T. Plourde, \PRB {\bf 79}, 174512 (2009).

\bibitem{GR} J. Gittleman and B. Rosenblum, \PRL {\bf 16}, 734 (1966).

\bibitem{Golo} M. Golosovsky, M. Tsindlekht, and D. Davidov, \SUST {\bf 9}, 1 (1996).

\bibitem{PompeoPRB} N. Pompeo and E. Silva, \PRB {\bf 78}, 094503 (2008).

\bibitem{Campbellpenetration} A. M. Campbell, \JPCSSP {\bf 2}, 1492 (1969).

\bibitem{rparameter1} J. Halbritter, \JS {\bf 8}, 691 (1995).

\bibitem{tsuchiya} Y. Tsuchiya, K. Iwaya, K. Kinoshita, T. Hanaguri, H. Kitano, A. Maeda, K. Shibata, T. Nishizaki and N. Kobayashi, \PRB {\bf 63}, 184517 (2001).

\bibitem{rparameter2} 
E. Silva, G. Ghigo, L. Gozzelino, C. Camerlingo and S.Sarti, \IJMPB {\bf 14}, 2822 (2000); 
A. V. Velichko, M. J. Lancaster, R. Chakalov, and F. Wellhofer, \PRB {\bf 65}, 104522 (2002); 
E. Silva, R. Marcon, R. Fastampa, M. Giura, S. Sarti and G. Ghigo, \JLTP {\bf 131}, 871 (2003); 
N. Pompeo, R. Rogai, E. Silva, A. Augieri, V. Galluzzi and G. Celentano, \JAP {\bf 105}, 013927 (2009).

\bibitem{kogan} V. G. Kogan, \PRB {\bf 24}, 1572 (1981); R. A. Klemm and J. R. Clem, \PRB {\bf 21}, 1868 (1980).

\bibitem{CCisoblique} M. W. Coffey and J. R. Clem, \PRB {\bf 45}, 10527 (1992). 

\bibitem{CC11757} M. W. Coffey and J. R. Clem, \PRB {\bf 46}, 11757 (1992).

\bibitem{brandtone} H. Brandt, \RPP {\bf 58}, 1465 (1995).

\bibitem{BGL} G. Blatter, V. B. Geshkenbein, and A. I. Larkin, \PRL {\bf 68}, 875 (1992).

\bibitem{blatterone} G. Blatter, M. V. Feigel'man, V. B. Geshkenbein, A. I. Larkin, and V. M. Vinokur, \RMP {\bf 66}, 1377 (1997).

\bibitem{Note1} Using a GR model for the interpretation of the data, the same applies to the \protect{$r$} parameter.

\bibitem{tinkham} M. Tinkham, {\it Introduction to Superconductivity, 2nd Ed.} (McGraw-Hill, United States of America, 1996).

\bibitem{cylres} N. Pompeo, R. Marcon, and E. Silva, \JSNM {\bf 20}, 71 (2007).

\bibitem{perturbation} P J. Petersan and S. M. Anlage, \JAP {\bf 84}, 3392 (1998).

\bibitem{Corbino} J. C. Booth, D. H. Wu and S. M. Anlage, \RSI {\bf 65}, 2082 (1994); D. H. Wu, J. C. Booth and S. M. Anlage, \PRL {\bf 75}, 525 (1995); N. Tosoratti, R. Fastampa, M. Giura, V. Lenzi, S. Sarti and E. Silva, \IJMPB {\bf 14}, 2926 (2000); 
K. Torokhtii, N. Pompeo, C. Meneghini, C. Attanasio, C. Cirillo, E.A. Ilyina, S. Sarti and E. Silva, J. Supercond. Nov. Magn., DOI 10.1007/s10948-012-1795-7 (2012).

\bibitem{circurrent} E. Silva, A. Lezzerini, M. Lanucara, S. Sarti and R. Marcon, \MST {\bf 9}, 275 (1998).
\bibitem{circurrent2} N. Pompeo, R. Rogai, K. Torokhtii, A. Augieri, G. Celentano, V. Galluzzi and E. Silva, \PC {\bf 479}, 160 (2012).

\bibitem{Collin} R. E. Collin, {\it Foundation for Microwave Engineering, 2nd Ed.} (McGraw-Hill
International Editions, Singapore, 1992)

\bibitem{submittedPompeo} N. Pompeo, A. Augieri, K. Torokhtii, V. Galluzzi, G. Celentano and E. Silva, {\it Vortex matter dynamics in YBaCuO with BaZrO3 nanocolumns as revealed by dc and high-frequency experiments: role of anisotropy and directional pinning}, submitted for publication (2012), arXiv:1211.3311 [cond-mat.supr-con].

\bibitem{PlacaisPRB54} B. Pla\ifmmode \mbox{\c{c}}\else \c{c}\fi{}ais, P. Mathieu, Y. Simon, E. B. Sonin and K. B. Traito, \PRB {\bf 54}, 13083 (1996).

\bibitem{examplesZthin} 
E. Silva, M. Giura, R. Marcon, R. Fastampa, G. Balestrino, M. Marinelli and E. Milani, \PRB {\bf 45}, 12566 (1992); 
S. Revenaz, D. E. Oates, D. Labb\'e-Lavigne, G. Dresselhaus and M. S. Dresselhaus, \PRB {\bf 50}, 1178 (1994); 
I. S. Ghosh, L. F. Cohen and J. C. Gallop, \SUST {\bf 10}, 936 (1997); 
J. R. Powell, A. Porch, R. G. Humphreys, F. Wellh\"ofer, M. J. Lancaster and C. E. Gough, \PRB {\bf 57} 5474 (1998); 
D. Neri, E. Silva, S. Sarti, R. Marcon, M. Giura, R. Fastampa and N. Sparvieri, \PRB {\bf 58}, 14581 (1998); 
T. Banerjee, V. C. Bagwe, J. John, S. P. Pai, R. Pinto and D. Kanjilal, \PRB {\bf 69}, 104533 (2004); 
E. Silva, R.Marcon, S.Sarti, R.Fastampa, M.Giura, M. Boffa and A. M. Cucolo, \EPJB {\bf 37}, 277 (2004); 
D. Janju\ifmmode \check{s}\else \v{s}\fi{}evi\ifmmode \acute{c}\else \'{c}\fi{}, M. S. Grbi\ifmmode \acute{c}\else \'{c}\fi{}, M. Po\ifmmode \check{z}\else \v{z}\fi{}ek, A. Dul\ifmmode \check{c}\else \v{c}\fi{}i\ifmmode \acute{c}\else \'{c}\fi{}, D. Paar, B. Nebendahl†, T. Wagner, \PRB {\bf 74}, 10450 (2006).

\bibitem{examplesZ}
R. Marcon, R. Fastampa, M. Giura and E. Silva, \PRB {\bf 43}, 2940 (1991); 
E. Silva, R. Marcon and F. C. Matacotta, \PC {\bf 218}, 109 (1993); 
A. Narduzzo, M. S. Grbi\ifmmode \acute{c}\else \'{c}\fi{}, M. Po\ifmmode \check{z}\else \v{z}\fi{}ek, A. Dul\ifmmode \check{c}\else \v{c}\fi{}i\ifmmode \acute{c}\else \'{c}\fi{}, D. Paar, A. Kondrat, C. Hess, I. Hellmann, R. Klingeler, J. Werner, A. K\"ohler, G. Behr and B. B\"uchner, \PRB {\bf 78}, 012507 (2008); 
T. Okada, H. Takahashi, Y. Imai, K. Kitagawa, K. Matsubayashi, Y. Uwatoko and A. Maeda, \PRB {86}, 064516 (2012).

\bibitem{Bespalov} A. A. Bespalov, and A. S. Mel’nikov, \PRB {\bf 85}, 174502 (2012).

\bibitem{Zaniso} T. B. A. Senior, {\it Radio Sci.} {\bf 13}, 639 (1978).

\bibitem{thinfilm} S. Sridhar, \JAP {\bf 63}, 159 (1988).
 
\bibitem{thinfilm2} P. Hartemann, \IEEEas {\bf 2}, 228 (1992); E. Silva, M. Lanucara, and R. Marcon, \SUST {\bf 9}, 934 (1996).

\bibitem{substrateeffects} N. Klein, H. Chaloupka, G. M\"{u}ller, S. Orbach, H. Piel, B. Roas, L. Schultz, U. Klein and M. Peiniger, \JAP {\bf 67}, 6940 (1990); 
L. Drabeck, K. Holczer, G. Gr\"uner and D. J. Scalapino, \JAP {\bf 68}, 892 (1990); 
E. Silva, M. Lanucara and R. Marcon, \PC {\bf 276}, 84 (1997); 
N. T. Cherpak, A. I. Gubin and A. A. Lavrinovich, {\it Telecommunications and Radio Engineering} {\bf 55} 81 (2001); 
N. Pompeo, R. Marcon, L. M chin, E. Silva, \SUST {\bf 18}, 531 (2005); 
N. Pompeo, L. Muzzi, V. Galluzzi, R. Marcon, E. Silva, \SUST {\bf 20}, 1002 (2007).

\bibitem{Johnk} C. T. A. Johnk, {\it Engineering Electromagnetic Fields and Waves, 2nd Ed.} (John Wiley \& Sons Inc., USA, 1988).

\end{thebibliography}
\end{document}